\documentclass{article}

\usepackage[a4paper,margin=1in]{geometry}
\usepackage{times}
\usepackage{graphicx}
\usepackage{amsmath,amssymb}
\usepackage{url}
\usepackage{hyperref}
\usepackage{xcolor}
\usepackage{booktabs}
\usepackage{float}

\title{Advancing Scientific Knowledge Retrieval and Reuse with a Novel Digital Library for Machine-Readable Knowledge}
\author{
Hadi Ghaemi$^{1}$, Lauren Snyder$^{1}$, Markus Stocker$^{1}$ \\
{\small $^{1}$TIB - Leibniz Information Centre for Science and Technology} \\
{\small \texttt{Hadi.Ghaemi@tib.eu, Lauren.Snyder@tib.eu, Markus.Stocker@tib.eu}}
}
\date{}

\begin{document}
\maketitle

\begin{abstract}
Digital libraries for research, such as the ACM Digital Library or Semantic Scholar, do not enable the machine-supported, efficient reuse of scientific knowledge (e.g., in synthesis research). This is because these libraries are based on document-centric models with narrative text knowledge expressions that require manual or semi-automated knowledge extraction, structuring, and organization. We present ORKG reborn, an emerging digital library that supports finding, accessing, and reusing accurate, fine-grained, and reproducible machine-readable expressions of scientific knowledge that relate scientific statements and their supporting evidence in terms of data and code. The rich expressions of scientific knowledge are published as reborn (born-reusable) articles and provide novel possibilities for scientific knowledge retrieval, for instance by statistical methods, software packages, variables, or data matching specific constraints. We describe the proposed system and demonstrate its practical viability and potential for information retrieval in contrast to state-of-the-art digital libraries and document-centric scholarly communication using several published articles in research fields ranging from computer science to soil science. Our work underscores the enormous potential of scientific knowledge databases and a viable approach to their construction. 
\end{abstract}

\vspace{0.5em}
\noindent
\textbf{Comments:} Accepted for publication at SIGIR 2025.\\
This is the authors’ accepted version (preprint) prior to ACM formatting and copyediting.\\
The official version of record is available in the ACM Digital Library at:\\
https://doi.org/10.1145/3726302.3730134
\vspace{1em}

\section{Introduction}
Research is an iterative process in which researchers build on existing knowledge to create new knowledge. Synthesis research is a key tool for creating new knowledge, and relies on primary scientific literature (i.e., scientific findings that are published for the first time) as a data source. For example, systematic reviews and meta-analyses are based on the integration and analysis of data extracted from published literature~\cite{yu2022synthesizing,depraetere2021critical}.
The validity of such syntheses depends on the quality of the primary scientific literature that is being synthesized~\cite{depraetere2021critical}. To assess this quality, researchers need to understand the data and methods underlying the scientific statements made in the literature.

Digital libraries, such as the ACM Digital Library or similar, support access to scientific articles, but are document-centric and the knowledge they contain is not machine-readable, meaning that to reuse this knowledge, researchers must first extract it. Manual data extraction is time consuming and semi-automated approaches usually trade extraction accuracy for speed to enable discovery and access to related data, code, or other supplementary materials. Digital libraries often interlink with artefacts distributed on specialized platforms such as GitHub, Kaggle, or (domain-specific) data repositories. Nonetheless, the ability of machines to reliably process scientific knowledge expressed in heterogeneous formats and published as sets of distributed files continues to be very limited.

With ORKG reborn---accessible at \href{https://reborn.orkg.org}{\textcolor{blue}{reborn.orkg.org}}---we rethink the publication of and access to machine-readable scientific knowledge in digital libraries for research. Inspired by~\cite{hars2001designing}, we use a conceptual model that organizes scientific knowledge (specifically, research findings expressed in articles) as statements and supporting evidence, expressed as machine-readable data with formal syntax. The proposed system consists of three interconnected layers: 1) the data deposition and collection layer, which facilitates the harvesting of reborn article~\cite{stocker2024rethinking} data published on distributed data repositories; 2) the knowledge organization layer, which provides a centralized database for the management and retrieval of reborn article data; and 3) the presentation layer, which renders reborn article data as structured scientific knowledge in the form of statements and supporting evidence in a user-friendly Web-based interface.

\section{Related Work}
The academic expansion of the past decades has dramatically increased the volume of published articles, scientific knowledge, and related artefacts such as data and code.
The global scholarly infrastructure~\cite{borgman2010scholarship} is designed to provide reliable and comprehensive access to these artefacts~\cite{verma2023scholarly}. Traditional systems enable discovery and access with persistent identifiers and associated standardized metadata (information about articles, authors, and datasets). Some systems, such as Zenodo\footnote{https://zenodo.org} support the deposition and interlinking of data and code. With the exception of specialized systems (e.g., domain-specific data repositories) these systems are generally unable to support direct access to and retrieval of content, be it research data in data files or scientific knowledge expressed in articles~\cite{aryani2018research,burton2017scholix,sadeghi2017integration}.

A new generation of systems aims to enable access to the content of articles by supporting collaborative manual or semi-automated extraction of knowledge from articles and collaborative curation and organization of extracted knowledge. In addition to the Open Research Knowledge Graph (ORKG)~\cite{jaradeh2019open},
Hi-Knowledge\footnote{https://hi-knowledge.org}~\cite{jeschke2020hiknowledge} aims to enable novel presentation and access to scientific knowledge in invasion biology and urban ecology, with the vision of creating an interactive atlas of knowledge covering other research fields. Similar systems can be found in other domains, e.g., Machine Learning\footnote{https://paperswithcode.com}.
A fundamental limitation of these systems is their reliance on post-publication knowledge extraction from articles, which is time consuming and error prone when done manually and requires tradeoffs in accuracy when completed with (semi)-automated approaches.

\begin{figure}[h]
  \centering
  \includegraphics[width=1.00\linewidth]{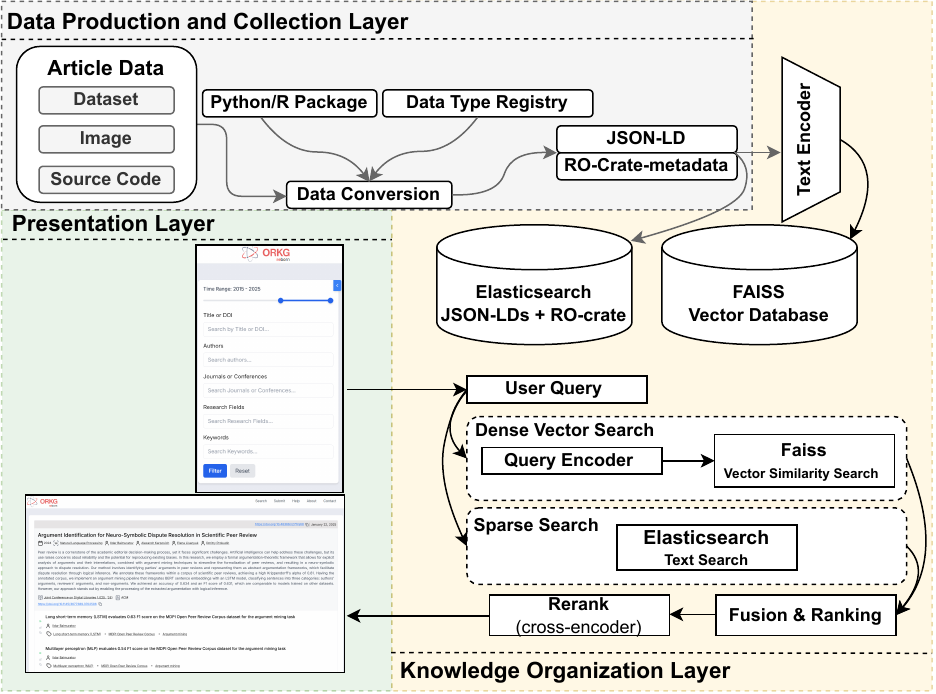}
  \caption{Proposed system architecture showing the three main layers: Data Deposition and Collection Layer, Knowledge Organization Layer, and Presentation Layer.}
  \label{fig:ARC}
\end{figure}

\section{Architectural Overview}
We now describe the architecture of the proposed system. As stated above, its aim is to publish machine-readable scientific statements and underlying evidence, with the aim of supporting synthesis research in particular. Toward this aim, it is necessary to provide reliable access not only to scientific findings, but also to related research data and source code. Figure \ref{fig:ARC} illustrates how we divide the overall architecture into three main layers: data deposition and collection layer, knowledge organization layer, and presentation layer.

\subsection{Data Deposition and Collection Layer}
To efficiently reuse research data, it is necessary to make data understandable to both humans and machines. This requires formalizing data types. We leverage a Data Type Registry (DTR) to manage the standardization and registration of, as well as access to, data types required to describe scientific statements and their underlying evidence. For unambiguous identification, each data type in the DTR is assigned a DOI as a unique, resolvable persistent identifier.

\begin{table}[h]
  \caption{Data types and their definition.}
  \label{tab:datatypes}
  \begin{tabular}{p{0.30\linewidth}p{0.60\linewidth}}
    \toprule
    Data type & Definition\\
    \midrule
    \href{https://doi.org/21.T11969/37182ecfb4474942e255}{\textcolor{blue}{Data Preprocessing}} & Prepare datasets for further analysis\\
    \href{	https://doi.org/21.T11969/5b66cb584b974b186f37}{\textcolor{blue}{Descriptive Statistics}} & Describe dataset characteristics\\
    \href{https://doi.org/21.T11969/5e782e67e70d0b2a022a}{\textcolor{blue}{Algorithm Evaluation}} & Algorithm performance evaluation for a specified task and dataset\\
    \href{https://doi.org/21.T11969/c6b413ba96ba477b5dca}{\textcolor{blue}{Multilevel Analysis}} & Multilevel data analysis with models including fixed and random effects\\
    \href{	https://doi.org/21.T11969/3f64a93eef69d721518f}{\textcolor{blue}{Correlation Analysis}} & Evaluate the strength and direction of the relationship between variables\\
    \href{	https://doi.org/21.T11969/b9335ce2c99ed87735a6}{\textcolor{blue}{Group Comparison}} & Compare the means of two or more groups\\
    \href{https://doi.org/21.T11969/286991b26f02d58ee490}{\textcolor{blue}{Regression Analysis}} & Explore the relationship between dependent variable(s) and independent variable(s)\\
    \href{https://doi.org/21.T11969/6e3e29ce3ba5a0b9abfe}{\textcolor{blue}{Class Prediction}} & Predict a categorical class label given input data\\
    \href{https://doi.org/21.T11969/c6e19df3b52ab8d855a9}{\textcolor{blue}{Class Discovery}} & Discover classes or clusters in unlabeled data\\
    \href{https://doi.org/21.T11969/437807f8d1a81b5138a3}{\textcolor{blue}{Factor Analysis}} & Discern latent factors in data\\
    \midrule
  \end{tabular}
\end{table}

\begin{figure}[h]
  \centering
  \includegraphics[width=1.0\linewidth]{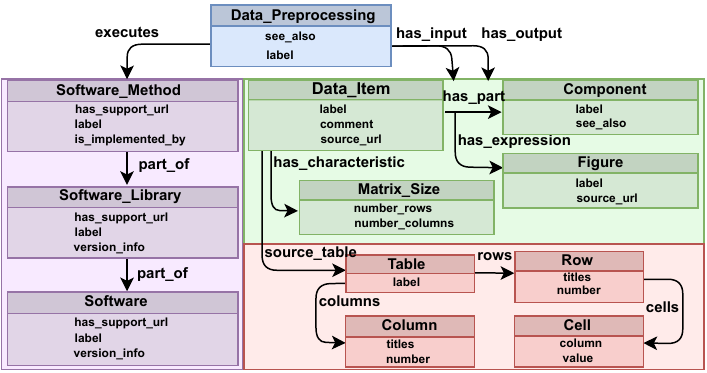}
  \caption{Diagram of the `Data Preprocessing' data type describing the executed procedure, the utilized input data, and produced output data.}
  \label{fig:DTR}
\end{figure}

Our current focus is on scientific statements where the underlying evidence is a statistical data analysis or set of data analyses, for which we have developed the data types listed in Table ~\ref{tab:datatypes}. Figure~\ref{fig:DTR} illustrates the data type for `Data Preprocessing', which captures information related to executed software methods, and input and output data items. For example, in this data type, software methods refer to functions in libraries that are part of (Python, R, etc.) software. Data items may be sourced from a tabular data structure or from a URL, are characterized by a matrix size and components, and may be expressed as a figure. For a full description of the `Data Preprocessing' data type, refer to \href{https://doi.org/21.T11969/37182ecfb4474942e255}{\textcolor{blue}{doi:21.T11969/37182ecfb4474942e255}}.

\begin{figure*}[htbp]
  \centering
  \includegraphics[width=\linewidth]{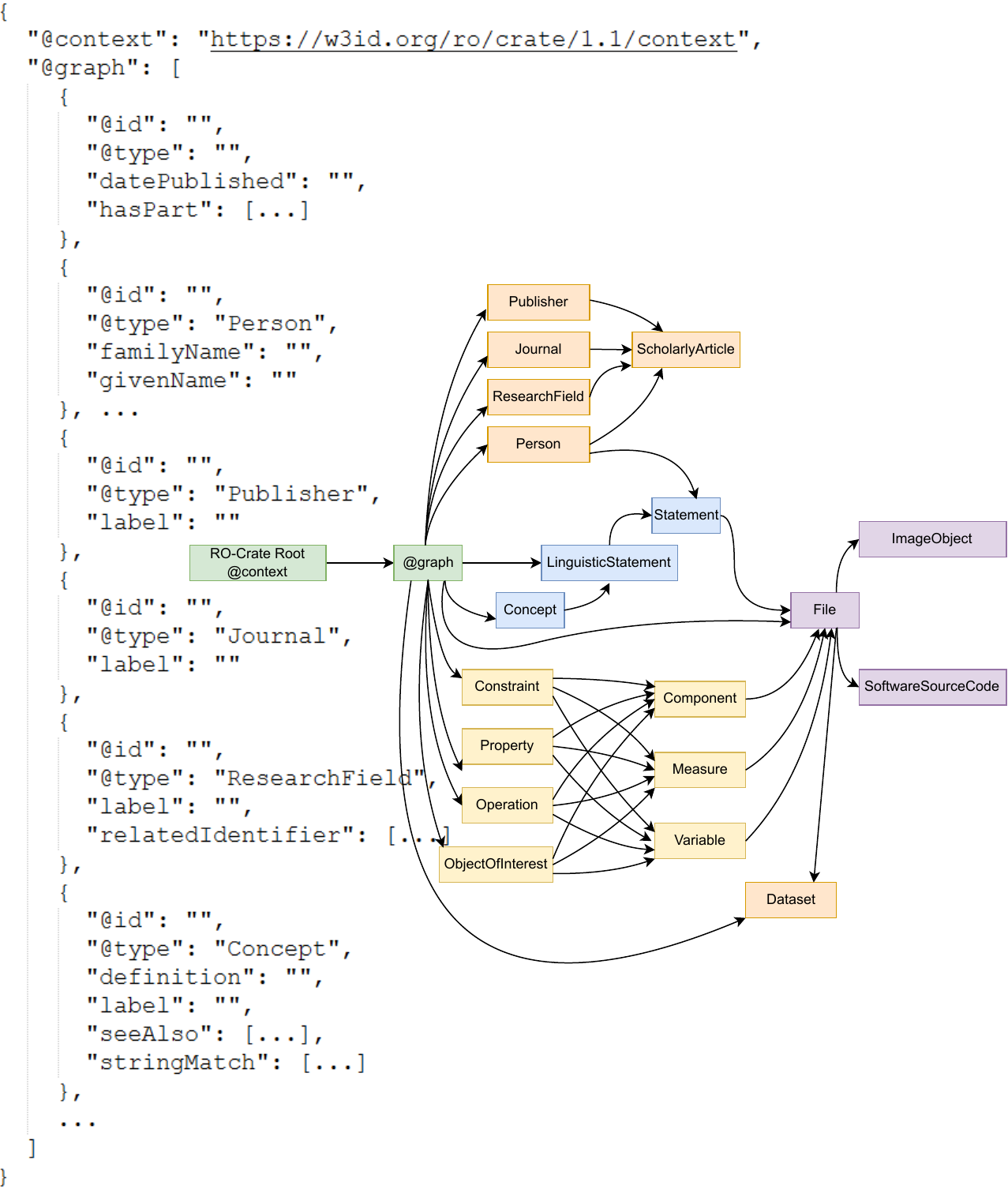}
  \caption{Overview of the RO-Crate metadata file structure.}
  \label{fig:R-crate}
\end{figure*}

The data describing scientific statements and their underlying evidence are deposited in JSON-LD format on a standard data repository. We leverage a CKAN-based system called the Leibniz Data Manager \cite{beer2022leibniz}. The proposed system deposits these data as RO-Crates \cite{Soiland_2022} with additional metadata about the article, authors, journals, publishers, statements, and data.
Figure \ref{fig:R-crate} shows the structure and graph of the RO-Crate metadata file. The root of this file consists of the \texttt{@context} and the \texttt{@graph}. The \texttt{@context} consists of the RO-Crate specification URL and the \texttt{@graph} includes different entity types described with their properties. Possible entity types include Dataset, Person, Publisher, Concept, Component, File, and Statement. By interlinking the data deposition with the original article in DOI metadata, we enable the discovery of this data by article DOI, and hence data collection in the proposed system.

\subsection{Knowledge Organization Layer}

The Knowledge Organization Layer represents the system's core data management and processing infrastructure. At its heart are two primary storage systems: Elasticsearch and the open source Faiss (Facebook AI Similarity Search) library~\cite{johnson2019billion}, each serving distinct but complementary purposes. Elasticsearch stores the structured JSON-LD representations of scientific statements and supporting evidence along with their associated RO-Crate metadata, thus providing robust document storage and traditional text search capabilities. The system uses a Text Encoder as a bridge between raw text and vector representations, processing incoming data and preparing data for semantic search capabilities.

% \begin{figure*}[H]
%   \centering
%   \includegraphics[width=1.0\linewidth]{Ro-crate.pdf}
%   \caption{Overview of the RO-Crate metadata file structure.}
%   \label{fig:R-crate}
% \end{figure*}

\subsection{Presentation Layer}
\begin{figure*}[ht]
  \centering
  \includegraphics[width=1\linewidth]{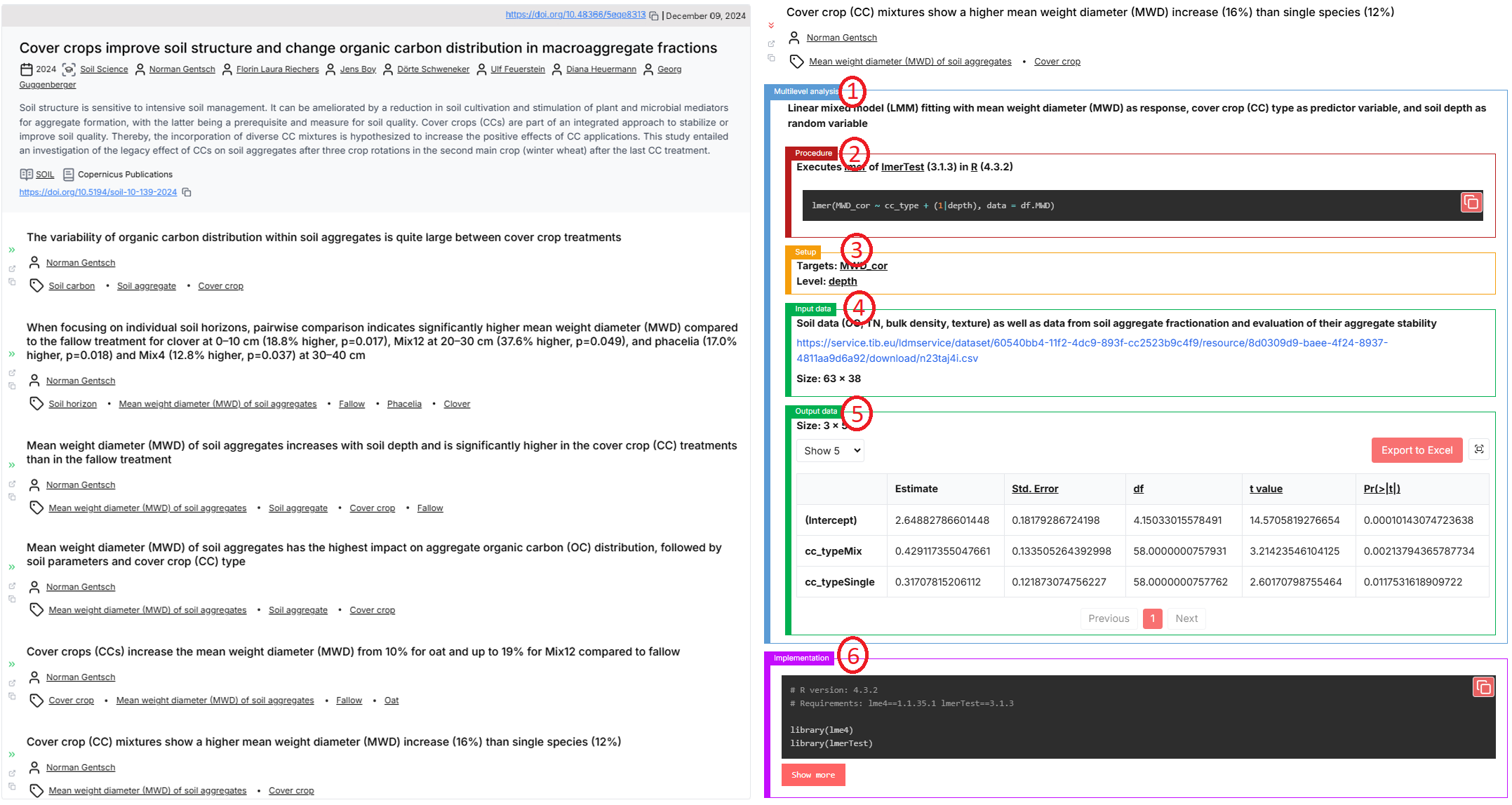}
  \caption{Scientific statements and supporting evidence as originally published by Gentsch et al.~\cite{gentsch2024cover} presented here as a reborn article accessible in the ORKG reborn digital library. (left) A reborn article presenting the original research findings as structured scientific statements and supporting evidence (\href{https://doi.org/10.48366/5eqe8313}{doi:10.48366/5eqe8313}). (right) Display of a scientific statement and supporting evidence in terms of a data analysis described by the executed procedure, utilized input data, produced output data, and full implementation in source code.}
  \label{fig:statment}
\end{figure*}

The Presentation Layer implements a Web-based user interface and supports interacting with the scientific knowledge database as well as knowledge retrieval and reuse. Figure \ref{fig:statment} illustrates how ORKG reborn presents a reborn article in terms of scientific statements (from here on referred to simply as statements) with supporting evidence in a structured manner that is intuitive for users. Each reborn article is presented with essential bibliographic metadata (title, authors, abstract, journal, DOI to the original PDF article) and a list of scientific statements published in the original work. Each reborn article is also assigned its own DOI, distinct from the original publication. Statements are annotated with concepts, which are described and utilized in the search interface. When statements are selected, the underlying evidence is displayed (Figure \ref{fig:statment}). 
Supporting evidence for statements is composed of different sections. In Figure \ref{fig:statment}, Section 1 describes the data analysis and its parts, e.g., a descriptive statistic. For each data analysis part, the system describes (a) the executed procedure (Section 2) in terms of the executed function of a specific package in Python or R languages; (b) the utilized input data (Section 4) and produced output data (Section 5), possibly associated with figures; (c) essential components such as target variables, also known as response variables (Section 3); as well as the full implementation of the data analysis in Python or R code (Section 6). As a result, the research methods and results are transparent, reproducible, and machine-readable, enabling other researchers to more easily confirm or reuse the published scientific knowledge.

\section{Implementation}
The presented system is deployed using Docker Compose and comprises a backend, a frontend, and infrastructure containers. The backend is implemented as a Web service in Python, and the frontend is implemented in Next.js. The textual information is indexed using Elasticsearch. 

As suggested in Figure~\ref{fig:ARC}, the user interface features a search interface that handles dense and sparse natural language queries. The dense search path uses the Query Encoder (powered by all-MiniLM-L6-v2~\footnote{https://huggingface.co/cross-encoder/ms-marco-MiniLM-L-6-v2}) to convert the user's query into a vector representation, which is then used by Faiss for vector similarity search. Concurrently, the keyword search path utilizes Elasticsearch's text search capabilities to find relevant documents based on traditional keyword matching algorithms. The results from both search paths are then combined through the Fusion \& Ranking component, which applies intelligent weighting to blend the results from both approaches. This hybrid approach ensures that users receive comprehensive results that capture both semantic relevance and keyword precision. The presentation layer then receives these ranked results and displays them in a clear and organized manner.

We utilize Faiss over Elasticsearch’s dense vector retrieval because of its high performance, GPU acceleration, and flexible index options, which optimize our semantic search for large-scale, low-latency requirements. 
FlatIndex supports the search for the exact nearest neighbor. As data increases, we can use approximation techniques to improve scalability. For instance, Hierarchical Navigable Small World (HNSW)~\cite{malkov2018efficient} finds nearest neighbors based on graphs, Inverted File Index (IVF)~\cite{sivic2003video} partitions vector spaces, and Product Quantization (PQ)~\cite{jegou2010product} compresses vectors to reduce memory usage.

For sentence embedding, we use the allmpnet-base-v2\footnote{https://huggingface.co/sentence-transformers/all-mpnet-base-v2} model, based on MPNet~\cite{song2020mpnet} and BERT~\cite{devlin-etal-2019-bert}. Following the embedding process, Faiss serves as the vector database, optimized for storing and retrieving high-dimensional vector embeddings. Complementing this vector-based approach, Elasticsearch provides robust full-text search. The integration of Faiss for vector storage and Elasticsearch for document storage establishes a comprehensive system that effectively handles both semantic and keyword-based queries. Finally, to perform the search, we implement a weighted score fusion methodology that integrates Elasticsearch and Faiss results. Furthermore, to enhance result quality, we use a cross-encoder (cross-encoder/ms-marco-MiniLM-L-6-v2) for re-ranking the top-10 aggregated results.

\section{Conclusions and Future Work}
This paper is a first, concise presentation of ORKG reborn, an emerging digital library that advances the machine-based retrieval and reuse of scientific knowledge. ORKG reborn is fueled by an initiative that advocates transitioning from born-digital, unstructured expressions of scientific knowledge in the form of traditional PDF articles to born-reusable and high-quality expressions of scientific knowledge that are produced machine-readable from the outset. By adopting a pre-publication approach, aka reborn articles, the proposed system and initiative distinguish themselves from classical approaches centered around post-publication knowledge extraction from articles. 

Building on a three-layered architecture, ORKG reborn supports the collection of reborn article data published in distributed data repositories, the organization of such machine-readable data in a centralized scientific knowledge database, and user-friendly presentation of the data in a Web-interface that supports data search, access, and download as packaged data collections.

In future work, we will further advance system capabilities along several directions. First, we will broaden the supported scientific knowledge types. Of particular interest are mathematical statements and supporting proofs. Second, enabled by the ZIP-packaged download of selected statements, we will develop use case projects in synthesis research that leverage ORKG reborn for scientific knowledge integration and synthesis. Third, inspired by Papers with Code, we will implement in ORKG reborn the display of algorithm evaluations as leaderboards and extend such specialized visualizations to other scientific knowledge types.

%%
%% The acknowledgments section is defined using the "acks" environment
%% (and NOT an unnumbered section). This ensures the proper
%% identification of the section in the article metadata, and the
%% consistent spelling of the heading.
\section{Acknowledgments}
The authors gratefully acknowledge the contributions of several colleagues, in particular Olga Lezhnina, Lars Vogt, and Manuel Prinz. This work has been supported by the Leibniz-Lab ``Systemic Sustainability'' (GA LL-2024-SYSTAIN) funded by the Leibniz Association and the Horizon Europe project FAIR2Adapt (GA 101188256) funded by the European Union.

%%
%% The next two lines define the bibliography style to be used, and
%% the bibliography file.
\bibliographystyle{plain}
\bibliography{references}

%%
%% If your work has an appendix, this is the place to put it.
\appendix

\end{document}